# Structural and Magnetic Phase Transitions in NdCoAsO under High Pressures

**Walter Uhoya, Georgiy Tsoi and Yogesh K. Vohra**
Department of Physics, University of Alabama at Birmingham (UAB)
Birmingham, AL 35294, USA

**Michael A. McGuire, Athena S. Sefat, Brian C. Sales, and David Mandrus**
Oak Ridge National Laboratory (ORNL)
Oak Ridge, TN 37831, USA

**Samuel T. Weir**
Mail Stop L-041, Lawrence Livermore National Laboratory (LLNL)
Livermore, CA 94550, USA

We have investigated structural and magnetic phase transitions under high pressures in a quaternary rare earth transition metal arsenide oxide NdCoAsO compound that is isostructural to high temperature superconductor NdFeAsO. Four-probe electrical resistance measurements carried out in a designer diamond anvil cell show that the ferromagnetic Curie temperature and anti-ferromagnetic Neel temperature increase with an increase in pressure. High pressure x-ray diffraction studies using a synchrotron source show a structural phase transition from a tetragonal phase to a new crystallographic phase at a pressure of 23 GPa at 300 K. The NdCoAsO sample remained anti-ferromagnetic and non-superconducting to temperatures down to 10 K and to the highest pressure achieved in this experiment of 53 GPa. A P-T phase diagram for NdCoAsO is presented to a pressure of 53 GPa and low temperatures of 10 K.

The pressure variable has always played a pivotal role in the discovery and optimization of novel superconducting materials and in the study of magnetic and structural phase transitions in materials under high pressures and low temperatures. The high temperature superconductivity in a new class of iron-based layered compounds has received extensive attention recently because of the diversity of systems in which this phenomenon has been documented [1]. Iron-based layered compounds like REFeAsO (RE= trivalent rare earth metal), $AFe_2Se_2$ (A = divalent alkaline earth metal or divalent rare earth metal), BFeSe (B= alkali metal), and simple FeSe(Te) materials have exhibited superconductivity with transition temperature ($T_c$) in the range of 0-55 K [1]. The quaternary rare earth transition metal arsenide oxides RETAsO where RE= trivalent rare earth metals and T= Fe, Ru, and Co have been synthesized [2] and have gained recent attention due to their isostructural nature with iron-based superconductors REFeAsO. The magnetic properties of RECoAsO have been examined in a series of rare earth compounds and Curie temperature ($T_C$) for the ferromagnetic transition was observed in the range of 55 K to 75 K [3]. Similarly, a ferromagnetic-anti-ferromagnetic transition ($T_N$) was observed in the range of 15 K to 75 K for a series of RECoAsO compounds [3]. In particular, magnetic phase transitions have been studied in detail in the compound NdCoAsO at ambient pressure down to 1.4 K at ambient pressure [4]. In NdCoAsO, a ferromagnetic phase transition is observed with $T_C$ = 69 K that is followed by two anti-ferromagnetic phase transitions at $T_{N1}$ = 14 K and $T_{N2}$ = 3.5 K respectively. In the anti-ferromagnetic states, magnetic moments for the cobalt atoms are oriented along the *a*-axis of the tetragonal lattice, which are compensated by the Nd moments that are pointed in the opposite direction. In the anti-ferromagnetic ordered state at 1.4 K, the ordered moment on the cobalt-site is $m_{Co} = 0.37\ \mu_B$ and on Nd-site is $m_{Nd} = 1.30\ \mu_B$ where $\mu_B$ is a Bohr's magnetron [4]. The present studies is motivated by studying the effects of pressure variable on the magnetic phase transitions and document any structural phase transitions that can be induced by pressure in this material. Finally, the possible occurrence of high temperature superconductivity in the compressed state in NdCoAsO is another motivation for this research.

The polycrystalline samples of NdCoAsO samples were prepared by solid state reactions as described in reference [4]. Our x-ray diffraction studies revealed a tetragonal structure with lattice parameters a = 3.987 ± 0.002 Å and c = 8.316 ± 0.001 Å with axial ratio c/a = 2.093 at ambient temperature and pressure. The tetragonal crystal structure is identified as ZrCuSiAs type with space group P4/nmm with Nd atoms at (1/4, 1/4, z-Nd), Co atoms at (1/4, 3/4, 1/2), As atoms at (1/4, 1/4, z-As) and O atoms at (1/4, ¾, 0). The structural parameters have been obtained from Rietveld refinements with z-Nd = 0.142 and z-As = 0.650 [4]. The high pressure x-ray diffraction experiments were carried out at the beam-line 16-BM-D, HPCAT, Advanced Photon Source, Argonne National Laboratory. An angle dispersive technique with an image-plate area detector was employed using a x-ray wavelength λ = 0.3757 Å. We employed eight probe designer diamond anvils [5, 6] in high pressure four-probe electrical resistance measurements on the NdCoAsO compound. The eight tungsten microprobes are encapsulated in a homoepitaxial diamond film and are exposed only near the tip of the diamond to make contact with the NdCoAsO sample at high pressure. Two electrical leads are used to set constant current through the sample and the two additional leads are used to monitor the voltage across the sample. The pressure was monitored by the ruby fluorescence technique and care was taken to carefully calibrate ruby $R_1$ emission to low temperature of 10 K as described in an earlier publication [7].

Fig. 1 shows the measured four-probe electrical resistance of NdCoAsO sample as a function of temperature between 10 K and 50 K and at various pressures between ambient pressure to 53 GPa. All the measured resistance curves have been normalized to the resistance value at 50 K to compare the resistance minimum observed at low temperatures. The ordering of magnetic moments gives rise to addition scattering of electrons and gives rise to an increase in the electrical resistance of the sample. The anti-ferromagnetic transition is marked by a minimum in electrical resistance at low temperature and is observed at 14 K at ambient pressure [4]. The minimum is marked by arrows or Neel temperature $T_{N1}$ in Fig. 1 and is observed to shift to higher temperature as the pressure is increased to 53 GPa. The electrical resistance minimum also becomes less pronounced as the pressure is increased, however, the presence of the minimum at the highest pressure indicated that material remains anti-ferromagnetic to the highest pressure of 53 GPa in the present experiments. It should be added that the second Neel temperature $T_{N2}$ of 3.5 K is below our low temperature limit of 10 K in the present series of experiments and we did not monitor $T_{N2}$ as a function of pressure. The measured Neel temperature $T_{N1}$ at various pressures can be fitted by the following equation:

$$T_{N1}(\text{in Kelvin}) = -\,0.0026 \cdot P^2 + 0.37 \cdot P + 18.7 \quad (0.8 < P < 53\ \text{GPa}) \qquad (1)$$

The electrical resistance measurements in an extended temperature range of 10 K to 100 K are plotted in Fig. 2 where a ferrmomagnetic Curie temperature ($T_c$) is seen as a change in slope of electrical resistance with temperature. The measured value of $T_C$ at ambient pressure is 69 K as measured in an earlier study [4]. The Curie temperature ($T_c$) is observed to increase with increasing pressure between ambient conditions to high pressures of 12.9 GPa. At higher pressures above 12.9 GPa, no infection is seen in the electrical resistance data so the location of the ferromagnetic transition is not established at ultra high pressures. The measured $T_C$ has been fitted to the following equation.

$$T_C(P) = 0.31 \cdot P + 75.0 \quad (0.8 < P < 13\ \text{GPa}) \qquad (2)$$

The high pressure x-ray diffraction experiments were carried out at the beam-line 16-BM-D, HPCAT, Advanced Photon Source, Argonne National Laboratory. An angle dispersive technique with an image-plate area detector was employed using a x-ray wavelength $\lambda = 0.3875$ Å. An internal copper pressure marker with a known equation of state [8] was utilized in our high pressure x-ray diffraction experiments and x-ray diffraction spectrums were recorded to 30 GPa at ambient temperature.

Fig.3 shows the integrated x-ray diffraction profiles for the NdCoAsO sample along with the copper pressure standard on increasing pressure at ambient temperature. The sample diffraction peaks in Fig. 3 are labeled by (hkl) values corresponding to a tetragonal phase while the copper pressure marker diffraction peaks are labeled according to a face-centered cubic structure. The sample is in the tetragonal phase at 0.3 GPa with lattice parameters a= 3.988 ± 0.003 Å and c= 8.321 ± 0.003 Å with c/a = 2.086 (Fig. 3 (a)). This tetragonal phase is retained to the highest pressure of 23 GPa. A representative integrated x-ray diffraction pattern in the intermediate pressure range is shown in Fig. 3 (b) at 11.3 GPa with lattice parameter a= 3.903 ± 0.003 Å, c= 7.979 ± 0.013 with c/a = 2.044. The spectrum at 25.8 GPa (Fig. 3(c)) is in the transformed phase as it shows additional diffraction peaks coexisting with the diffraction peaks of tetragonal phase. These additional diffraction peaks are marked by an asterisk in Fig. 3(c) indicating a structural phase transformation. Since the mixture of tetragonal phase and the new phase persists in the x-ray diffraction pattern to the highest pressure of 30 GPa, no attempt was made in the refine the crystal structure as single phase diffraction patterns are needed.

Further evidence of a phase transformation is observed in the plot of axial ratio (c/a) as a function of pressure (Fig. 4). In the low pressure range, c/a decreases rapidly with increasing pressure as the Co-As tetrahedral layers come closer on compression (Fig.4). This observed rapid decrease in (c/a) ratio with pressure appears to be a common phenomenon observed in other Iron-based layered superconductors under high pressure [9]. This sharp decrease in c/a ratio with increasing pressure abruptly ends at the phase transition at 23 GPa as indicated by an arrow in Fig. 4. The measured Pressure-Volume (P-V) curve or equation of state for the tetragonal phase of NdCoAsO is shown in Fig.5. The equation of state was fitted to the following Birch-Murnaghan equation of state (EoS) [10] shown in equation (3). The Birch Murnaghan equation was also fitted to the available data on copper pressure standard and details on the fit are provided in reference [8].

$$P = 3B_0 f_E (1 + 2f_E)^{5/2} \left\{ 1 + \frac{3}{2}(B' - 4) f_E \right\} \qquad (3)$$

Where $B_o$ is the bulk modulus, B' is the first derivative of bulk modulus at ambient pressure, and $V_0$ is the ambient pressure volume. Our measured value of ambient pressure unit-cell volume for NdCoAsO is 132.21 Å$^3$. The parameter $f_E$ is related to volume compression and is described below.

$$f_E = \frac{\left[\left(\frac{V_o}{V}\right)^{2/3} - 1\right]}{2}$$

The P-V data on NdCoAsO to 23.6 GPa at ambient temperature is shown in Fig. 5 and can be fitted to the Birch-Murnaghan equation (1) with Bulk Modulus ($B_0$) = 111.2 GPa and its pressure derivative B' = 2.357. This fit is also displayed in Fig. 5.

The measured phase-diagram (P-T diagram) for NsCoAsO is shown in Fig. 6. The application of external high pressure is shown to increase the anti-ferromagnetic transition temperature ($T_{N1}$). The pressure variation of the ferromagnetic transition temperature ($T_C$) is also shown in Fig. 6. The lower Neel Temperature ($T_{N2}$) was below our low temperature limit of 10 K and was not monitored in the present series of experiments and is indicated by a dashed line in Fig. 6. The measured transition pressure for the structural transition is 23 GPa at 300 K. Since the structural transition indicated by $T_s$ in Fig. 6 has not been studied at low temperatures, the phase boundary for $T_S$ is only indicated by a dashed line. The solid curves in Fig. 6 are the fits to the experimental data given by equations (1) and (2).

In conclusion, the anti-ferromagnetic phase in the quaternary rare earth transition metal arsenide oxide NdCoAsO was found to be stable to the ultra high pressure of 53 GPa and temperatures to 10 K. Since the suppression of anti-ferromagnetic phase is correlated to the appearance of a superconducting phase in Iron-based layered compounds, as a consequence no superconducting state was discovered in NdCoAsO till 53 GPa and 10 K. The Curie Temperature ($T_C$) and the upper Neel Temperature ($T_{N1}$) were both observed to increase with increasing pressure. The ambient pressure tetragonal phase transformed to a new crystal structure above 23 GPa at 300 K with the appearance of additional diffraction peaks. The crystal structure of the new phase was not refined due to the persistence of phase mixtures to the highest pressure of 30 GPa. Further low temperature-high pressure crystal structure studies would be needed to complete the phase diagram of quaternary rare earth transition metal arsenide oxide NdCoAsO.

**ACKNOWLEDGMENT**

Walter Uhoya acknowledges support from the Carnegie/Department of Energy (DOE) Alliance Center (CDAC) under Grant No. DE-FC52-08NA28554. Work at Oak Ridge National Laboratory is supported by the DOE-Basic Energy Sciences, Division of Materials Sciences and Engineering. Portions of this work were performed at HPCAT (Sector 16), Advanced Photon Source (APS), Argonne National Laboratory.

Figure Captions:

Fig. 1: The temperature variation of the electrical resistance normalized to its value at 50 K is shown for NdCoAsO at various pressures. The location of Neel Temperature ($T_{N1}$) is indicated by a minimum in the electrical resistance with respect to temperature (marked by arrows). The Neel Temperature is observed to increase with increasing pressure to 53 GPa.

Fig. 2: The measured variation of electrical resistance with temperature is plotted in an extended temperature range of 10 K to 100 K for NdCoAsO at various pressures. The Curie temperature ($T_c$) is indicated by a slight change in slope of the electrical resistance plots with temperature as marked by arrows. The Curie temperature is observed to increase with increasing pressures to 13 GPa and was not detected above this pressure.

Fig. 3: The integrated x-ray diffraction profiles for NdCoAsO sample and copper pressure standard at various pressures at T = 300 K recorded with x-ray wavelength $\lambda$ = 0.3757 Å. The (hkl) Miller indices are indicated for NdCoAsO tetragonal phase and face-centered cubic phase for copper. (a) Sample is in the ambient pressure tetragonal phase at 0.3 GPa, (b) sample is in the tetragonal phase at a higher pressure of 11.5 GPa, and (c) x-ray spectrum after phase transformation in sample at 25.8 GPa with three extra diffraction peaks marked by an asterisk.

Fig. 4: The measured axial ratio (c/a) as a function of pressure for the tetragonal phase. The c/a ratio shows a pronounced decrease with increasing pressure indicating a strong tendency for tetrahedral bonded Co-As layer to approach each other. This decrease comes to an end on transformation to a new phase at 23 GPa.

Fig. 5: The measured volume compression ($V/V_0$) versus pressure for the tetragonal phase of NdCoAsO to a pressure of 23.6 GPa. The solid curve is a fit to the Birch Murnaghan equation of state with parameters described in the text.

Fig. 6: The experimental Temperature-Pressure (T-P) diagram for NdCoAsO based on the present series of experiments. The pressure variation of Curie temperature ($T_C$), upper Neel

temperature ($T_{N1}$) is shown. The structural phase transformation ($T_S$) at room temperature is indicated but the phase boundary at low temperature is not yet established.

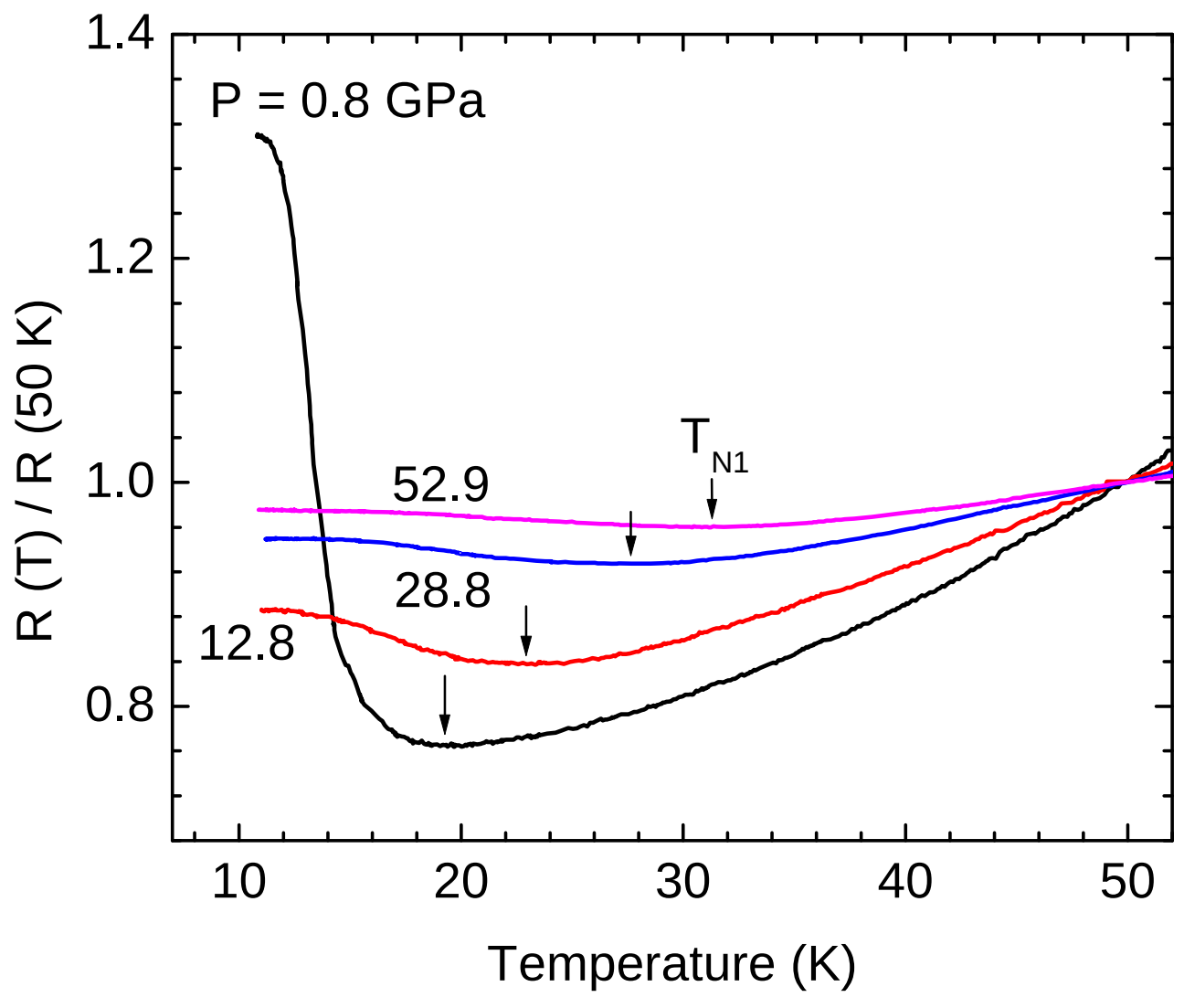

Figure 1

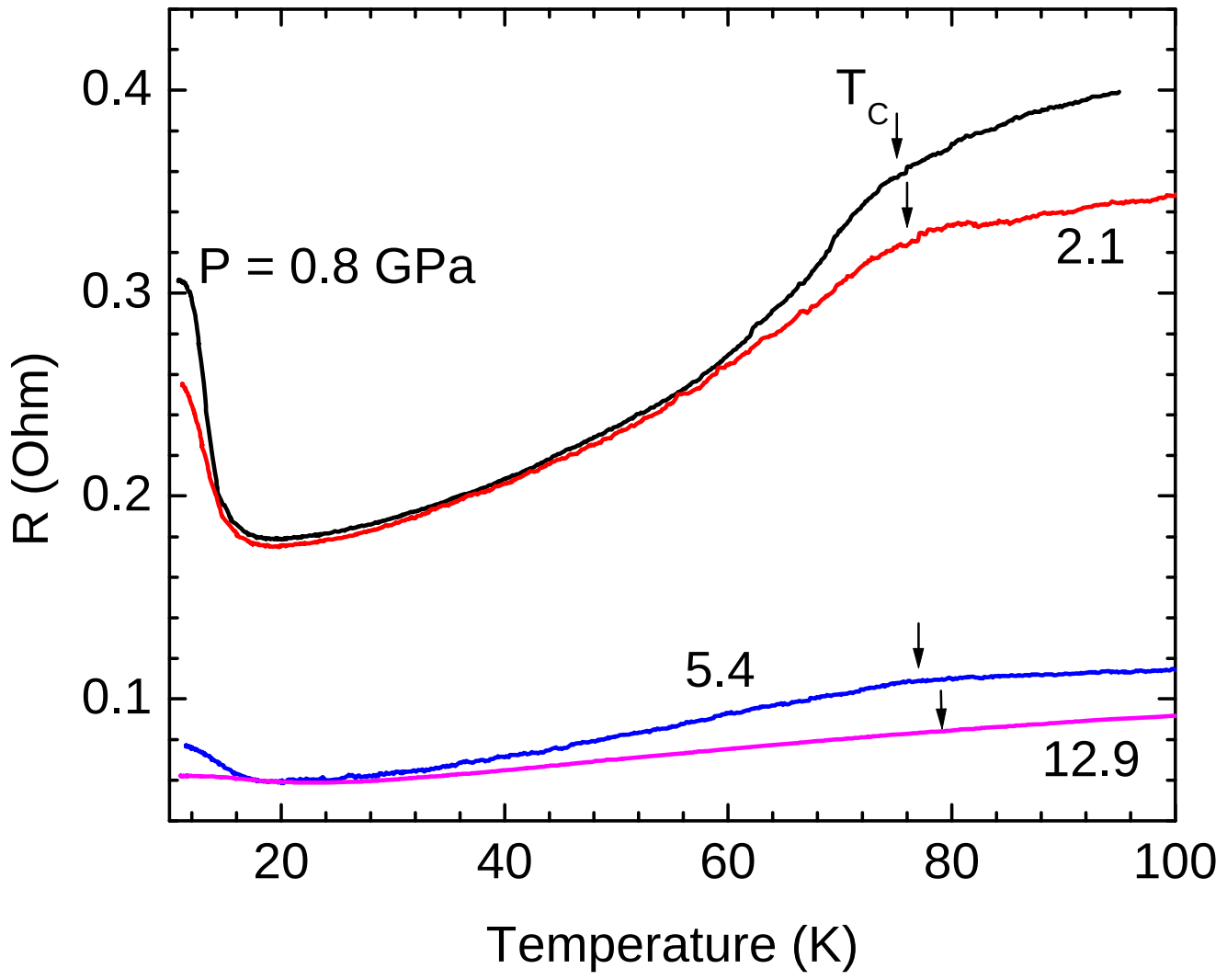

Figure 2

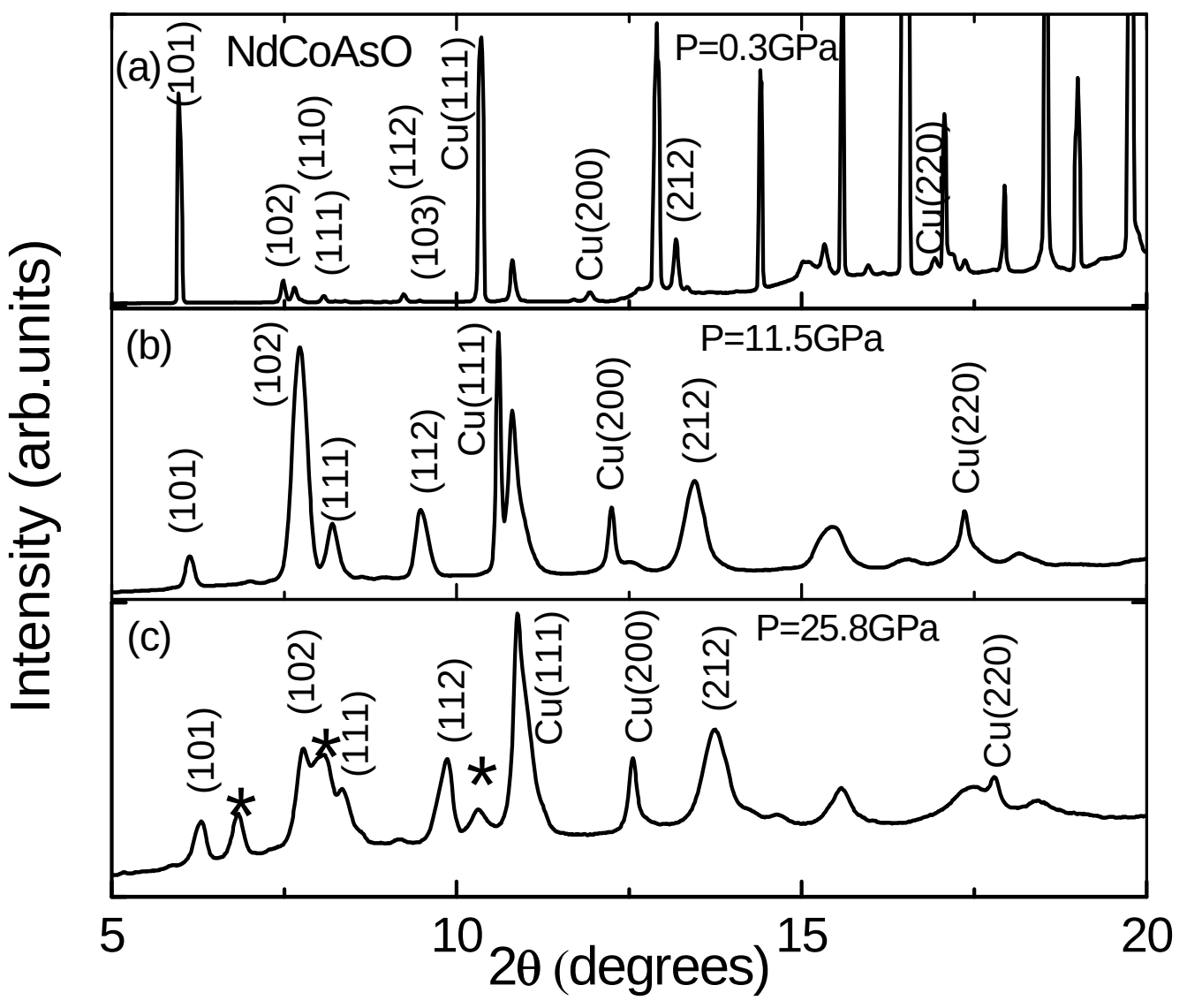

Figure 3

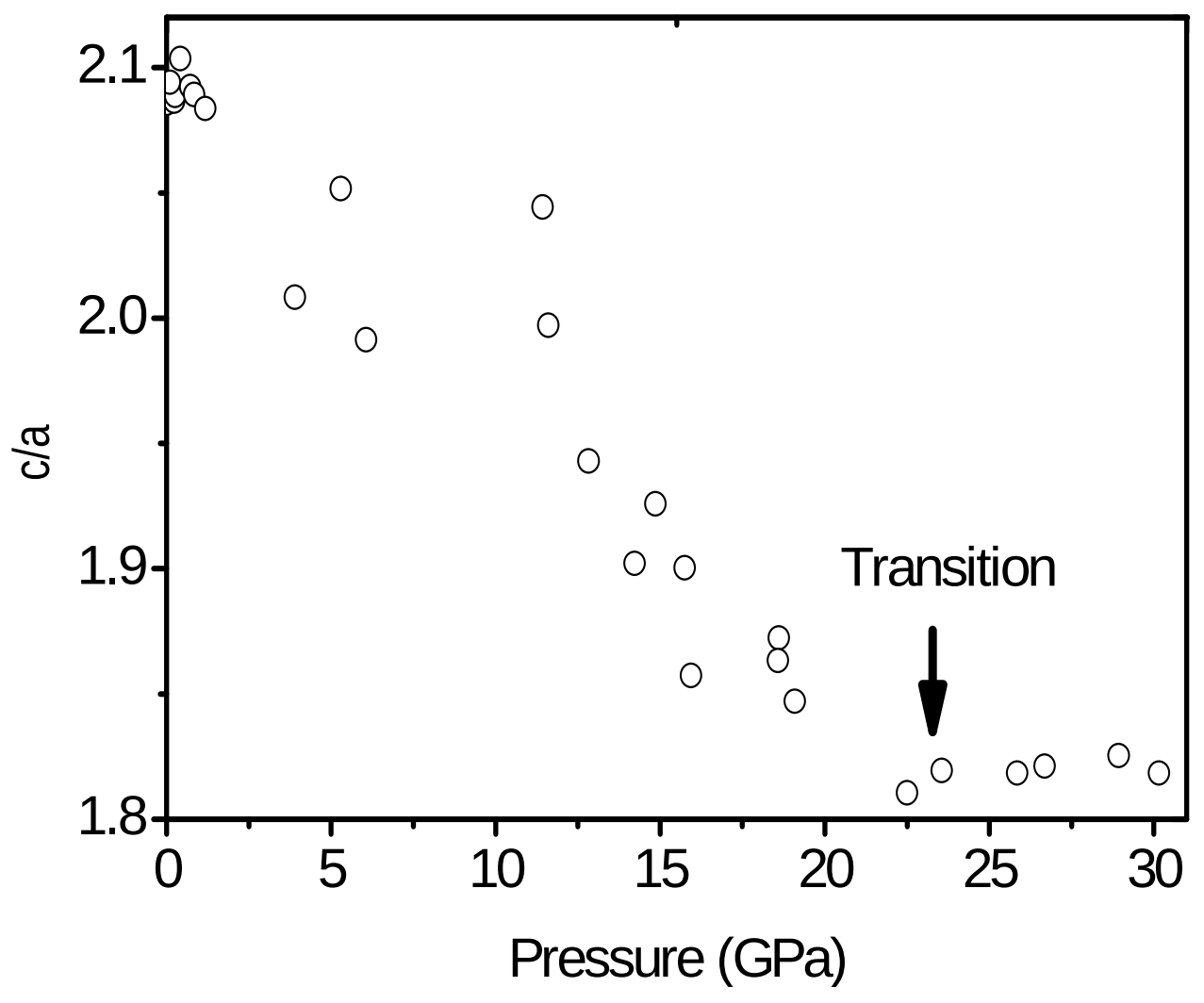

Figure 4

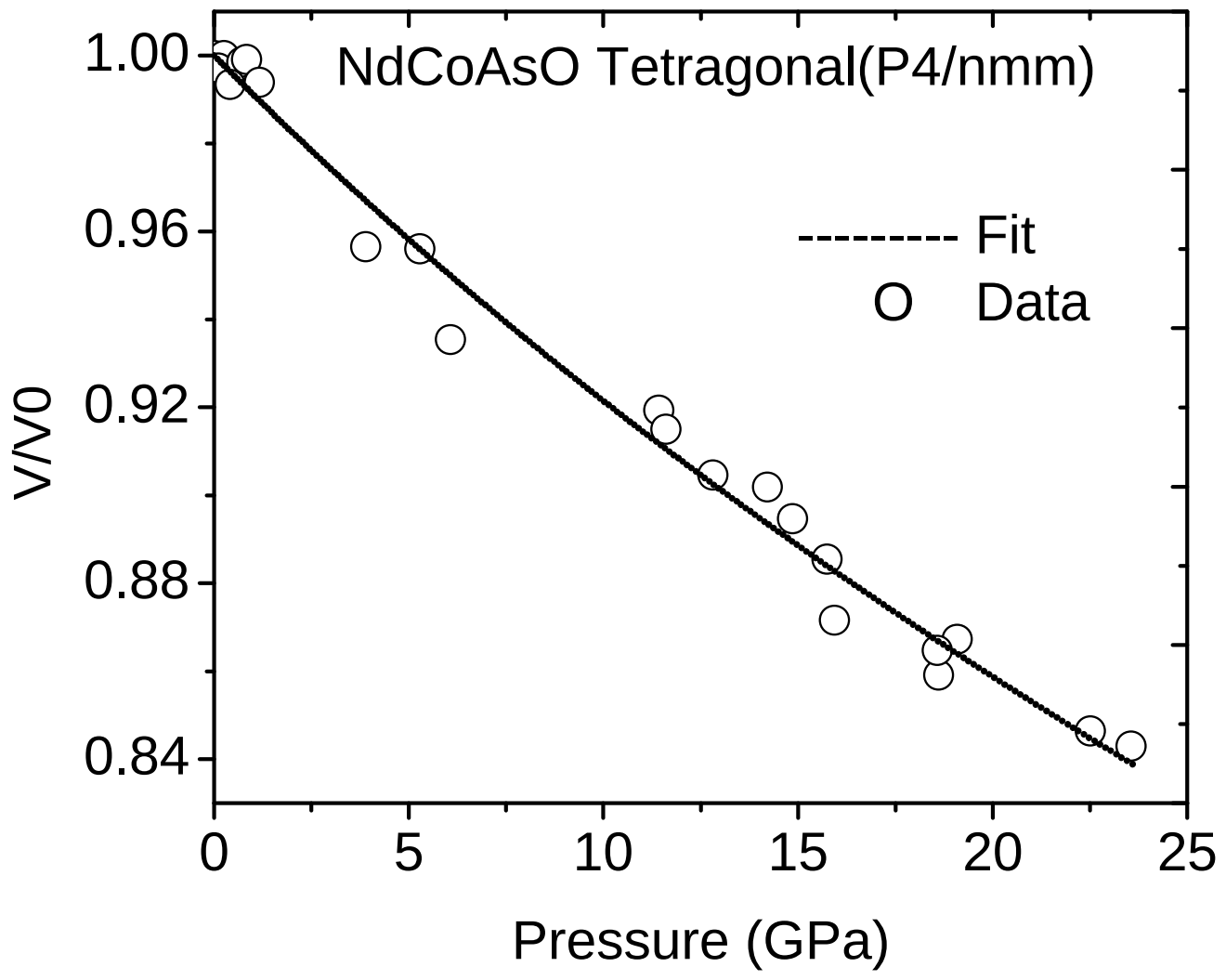

Figure 5

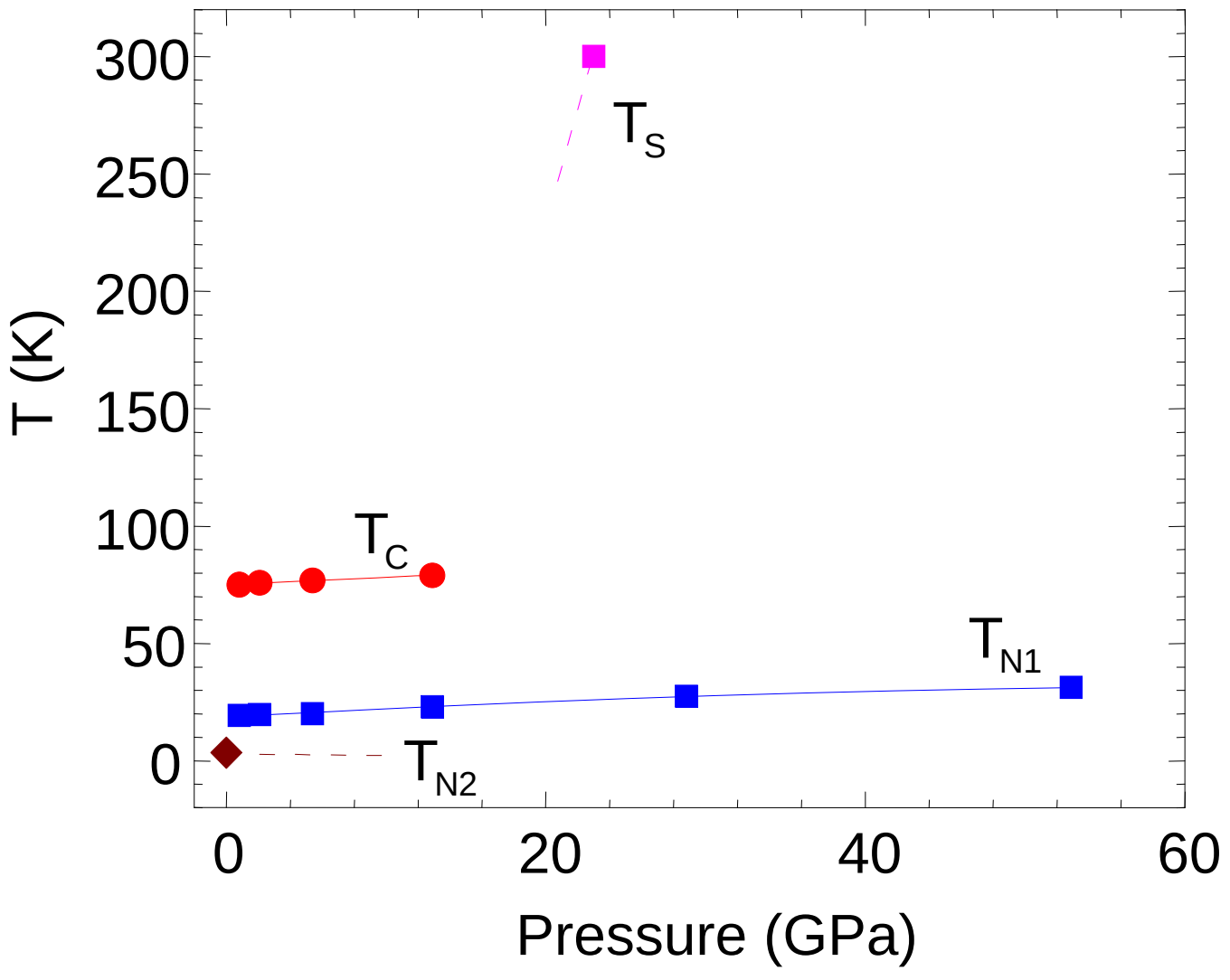

Figure 6